# Allium Vegetables Intake and Digestive System Cancer Risk: A Study Based on Mendelian Randomization, Network Pharmacology and Molecular Docking

Shuhao Li [1,2], Jingwen Lou [1,2], Yelina Mulatihan [1,2], Yuhang Xiong [1,2], Yao Li[1] and Qi Xu[1]*

[1] Department of Immunology, School of Basic Medical Sciences, Xinjiang Medical University，Urumqi 830000，Xinjiang，China
[2] The Department of Clinical Medicine，Xinjiang Medical University，Urumqi 830000，Xinjiang，China
* Correspondence: xjmyxq@xjmu.edu.cn

**Abstract:** Background: Allium vegetables (garlic and onion) are one of the flavorings in people's daily diets. Observational studies suggest that intake of allium vegetables may be correlated with a lower incidence of digestive system cancers. However, the existence of a causal relationship is still controversial due to confounding factors and reverse causation. Therefore, we explored the causal relationship between intake of allium vegetables and digestive system cancers using Mendelian randomization approach.

Methods: First, we performed Mendelian randomization analyses using inverse variance weighting (IVW), weighted median, and MR-Egger approaches, and demonstrated the reliability of the results in the sensitivity step. Second, Multivariable Mendelian randomization was applied to adjust for smoking and alcohol consumption. Third, we explored the molecular mechanisms behind the positive results through network pharmacology and molecular docking methods.

Results: The study suggests that increased intake of garlic reduced gastric cancer risk. However, onion intake was not statistically associated with digestive system cancer.

Conclusion: Garlic may have a protective effect against gastric cancer.

**Keywords:** Allium vegetables; digestive system cancer; Mendelian randomization; network pharmacology; molecular docking

## 1. Introduction

Digestive system cancers, mainly esophageal, gastric, liver, pancreatic, and colorectal cancers, are one of the most important causes of death in the global population, with about 4.8 million new cases and 3.1 million deaths annually [1]. Digestive system cancers have become one of the most critical public health problems [2]. Allium is the largest and most important representative genus of plants in the allium family Alliaceae and is widely distributed in the northern hemisphere. Onion and garlic belong to the allium and are among the oldest cultivated plants used for food and medicine [3]. The main components of garlic and other allium vegetables, for example, organosulfur compounds, have been shown to have anti-inflammatory, cardiovascular and obesity-preventing effects [4-6]. Studies have shown that when garlic is damaged or compressed, it produces allicin, which is unstable and can be further broken down into DAS, DADS, and DATS. Several recent studies have shown that allicin and its breakdown components have therapeutic effects on cancer, especially digestive system cancers [7-10]. In addition to the therapeutic effects of the active ingredients of allium vegetables on digestive system cancers, several epidemiological studies have been conducted in recent years on the intake of allium vegetables and the digestive system cancer risk [11-17]. However, the results of observational studies between onion and garlic intake and digestive system cancers risk lack consistency. European Prospective Investigation into Cancer and Nutrition (EPIC), a negative correlation exists between total intake of onions and garlic and the risk of developing gastric cancer [14]. However, after long-term follow-up and reanalysis, onion and garlic intake were found not to be correlated with the gastric cancer risk [11].

Traditional observational studies are vulnerable to recall bias, residual confounders, and reverse causation, which may confound results regarding the relationship between intake of allium vegetables and digestive system cancer risk. Mendelian randomization is an epidemiological statistical approach to assess the causal relationship between exposure and outcome using genetic variants as an instrumental variable (IVs) [18]. Since genetic variants are randomly assigned through meiosis, they are determined at the time of formation of the fertilized egg, therefore, not subject to reverse causation. Mendelian randomization methods maximize the exclusion of confounding factors and do not have the limitations of traditional epidemiological analysis [19]. The method has the same effect as randomized controlled trials [20].

In this study, we investigated the causal relationship between the intake of allium vegetables (garlic and onion) and cancers of the digestive system using Mendelian randomization (MR). In addition, we explored the molecular mechanisms of exposures and outcomes where a causal relationship exists through network pharmacology and molecular docking techniques. We expect that this study will provide new insights into the relationship between Allium vegetables and digestive system cancers, elucidating the potential causality and molecular mechanisms between the two.

**2.Materials and Methods**

*2.1. Study Design*

We reported our study following the STROBE-MR statement (Reporting of Observational Studies Using Mendelian Randomization to Enhance Epidemiology) [21]. The STROBE-MR checklist is provided in Supplementary Table 1. The research steps in this study were divided into three steps: First, we performed MR analysis between garlic, onion intake and the digestive system cancers. Secondly, we adjusted for smoking and alcohol consumption in Multivariable MR randomization analysis. Finally, we used network pharmacology and molecular docking approach to investigate the molecular mechanisms of garlic acting on gastric cancer. This study of Mendelian randomization analysis was conducted following the process illustrated in Figure 1.

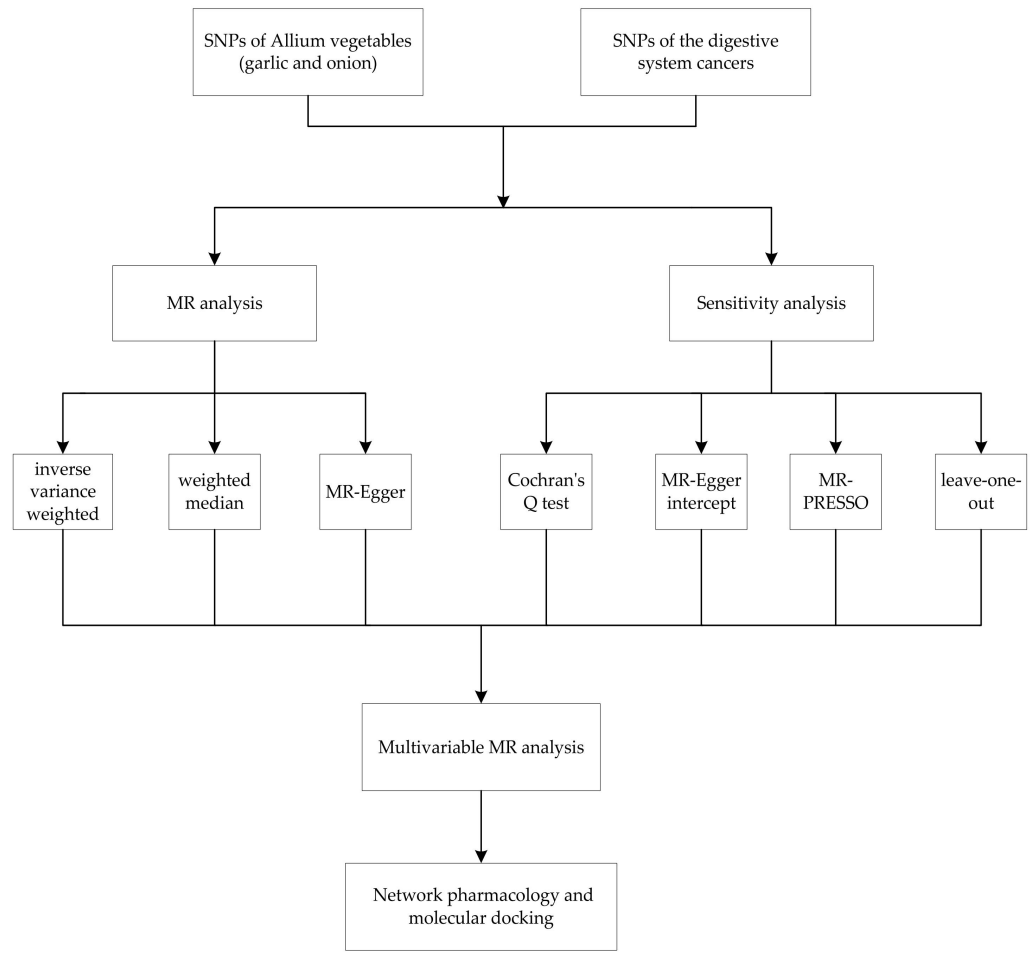

**Figure 1.** Flowchart of MR analysis for evaluating allium vegetables intake and risk of digestive system cancer; MR: Mendelian randomization.

## 2.2. Obtaining Instrumental Variables (IVs)

The selection of IVs fulfils three core assumptions: first, genetic instrumental variables (IVs) were strongly correlated with garlic and onion intake; second, IVs were not correlated with any of the factors influencing exposure and outcome; and third, IVs could only influence outcome by exposure [22]. We set a genome-wide significance level (P<5E-8) to obtain SNPs as instrumental variables and trimmed for linkage disequilibrium (LD) (kb=10,000, $R^2$<0.001). We set a genome-wide significance level (P<5E-8) to obtain SNPs as instrumental variables and trimmed for linkage disequilibrium (LD) (kb=10,000, $R^2<0.001$). However, based on the genome-wide threshold ($P<5E-8$) was not able to obtain SNPs with genome-wide significance for garlic and onion intake in European populations. For sub-sequent analyses, we modified the genome-wide threshold ($P<5E-06$) and calculated the F statistic to estimate the strength of IVs [23]. The weak instruments ($F<10$) will be removed, and we will not use the weak instruments in the MR analysis. Also, we used LDtrait [24] and the IEU QpenGWAS project (https://gwas.mrcieu.ac.uk/) to remove potential confounders in IVs, such as smoking, alcohol consumption, BMI, and so on. During the selection of instrumental variables, we also removed palindromic SNPs.

*2.3. GWAS data sources for exposure, outcome, covariates*

We acquired the required data from the publicly accessible Genome-Wide Association Study Statistics. Data on garlic and onion intake in the European populations were obtained from the UK Biobank, a large prospective cohort study that recruited 500,000 volunteers aged 40-69 years from England, Scotland, and Wales (94 percent of whom indicated their European origin) [25]. The information regarding the intake of allium vegetables was collected retrospectively through the Food Frequency Touchscreen Questionnaire. Garlic intake was based on the question, "How much garlic did you eat yesterday?" Participants could determine the frequency of garlic intake using the following options (quarter, half, 1, 2, 3+). Onion intake based on the same question, participants could determine the frequency of onion intake from the following options (quarter, half, 1, 2, 3+). If the answer was unrealistic, the submission was rejected. We obtained GWAS data on participants' garlic and onion intake as the exposure for this study.

We selected five common digestive system cancers (esophageal, gastric, liver, pancreatic, and colorectal) as the outcomes. GWAS data for esophageal, gastric, pancreatic and colorectal cancers were obtained from the European Bioinformatics Institute[26], while GWAS data for liver cancer were obtained from the FinnGen consortium R9 release data [27]. In the multivariable Mendelian randomization stage of the analysis, to use adjustments for smoking and alcohol consumption, we obtained GWAS data on smoking and alcohol consumption from IEU Open GWAS databases. This study was based on publicly available data. All dataset participants were European, and all data were collected with the participants' informed consent. Participants provided written informed consent, so no additional ethical approval or consent was required.

**Table 1.** GWAS data sources for exposure, outcome and covariates

| Trait | Ancestry | N | N cases | Number of SNPs | Data Source |
|---|---|---|---|---|---|
| Exposure | | | | | |
| Garlic intake | European | 649,49 | - | 9,851,867 | ukb-b-17223 |
| Onion intake | European | 649,49 | - | 9,851,867 | ukb-b-8124 |
| Outcome | | | | | |
| Esophageal cancer | European | 476,306 | 998 | 24,194,380 | GCST90018841 |
| Gastric cancer | European | 476,116 | 1029 | 24,188,662 | GCST90018849 |
| Liver cancer | European | 174,310 | 304 | 16,380,303 | finn-b-C3_LIVER_INTRAHEPATIC_BILE_DUCTS_EXALLC |
| Pancreatic cancer | European | 476,245 | 1196 | 24,195,229 | GCST90018893 |
| Colorectal cancer | European | 470,002 | 6581 | 24,182,361 | GCST90018808 |
| Covariate | | | | | |
| Ever vs never smoked | European | 74,035 | 419,69 | 2,455,847 | ieu-a-962 |
| Alcohol consumption | European | 295,40 | - | 7,914,327 | ieu-b-4833 |

*2.4. Mendelian randomization analysis*

In this study, we used the following methods for MR analysis of garlic, onion intake and digestive system cancers. We conducted the preliminary analyses by IVW, which is considered the most reliable outcome when all IVs are valid and can provide a reliable causal estimation without directed polymorphisms[28]. The weighted median method provides consistent effect estimates when above 50% of the IVs are valid [29]. Although all SNPs are ineffective, we can still use the Egger method to analyze them [30]. Combining these three Mendelian randomization analysis methods can provide more robust results for causal estimation.

*2.5. Sensitivity analysis*

In performing the sensitivity analysis phase, we used Cochran's Q test, MR-Egger intercept method, outlier (MR-PRESSO), and leave-one-out method. First, to evaluate the heterogeneity among individual SNPs, we computed Cochran's Q. Then, we used the intercept of the MR-Egger regression model to test whether the SNPs were horizontally pleiotropic. If p>0.05, the SNP is considered not to be horizontally pleiotropic. Finally, we detected the presence of outliers in the SNPs by MR-PRESSO [31]. If there are outliers, we need to remove the outliers and calculate the MR results of the remaining SNPs. Overall, we used the "leave-one-out" method in the sensitivity analysis stage, which is a method to observe whether the results are stable after removing the outliers by continuously removing the outliers and calculating the meta-effects of the remaining SNPs.

*2.6. Multivariable MR analysis*

We performed a Multivariable MR analysis. Multivariable MR analysis is a method to estimate multiple exposures and one outcome at the same time [32]. We further estimated the causal association between intake of allium vegetables and cancers of the digestive system after adjusting for smoking and alcohol consumption. MVMR still used inverse variance weighted, weighted median, MR-Egger approaches.

*2.7. Statistical Analysis*

The analyses and visualization of the results used in this study were performed using R statistical software (version 4.3.2) and the software packages "TwosampleMR",

"forestplot", "MRPRESSO" and "mendelanrandomization". For MR analysis and sensitivity analysis results, p<0.05 was regarded as statistically significant.

*2.8. Network pharmacology and molecular docking*

Based on the results of the MR analysis, we tentatively conclude that intake of garlic is a protective factor against gastric cancer. Therefore, we conducted a network pharmacology and molecular docking study on garlic and gastric cancer.

2.8.1. Acquisition of garlic active ingredients and related targets

The HERB database (http://herb.ac.cn/) was searched for "garlic" and the composition of garlic was obtained [33]. We screened the active ingredients based on relative molecular mass (MW) ⩽ 500, lipid-water partition coefficient (xlogp) ⩽ 5, number of hydrogen bond donors ⩽ 5, and number of hydrogen bond acceptors ⩽ 10. We obtained the SMILES numbers corresponding to the active ingredients of garlic from the PubChem database [34]. Subsequently, the SMILES number was used to predict the corresponding target of the active ingredient in Swiss Target Prediction (http://swisstargetprediction.ch/), with the screening criterion of "Homo sapiens", and duplicates were removed.

2.8.2. Acquisition of targets related to gastric cancer

The keywords "gastric cancer" was searched in Genecards [35] and OMIM [36], and the targets related to gastric cancer were obtained after removing duplicates. Subsequently, we extracted the intersection of garlic and gastric cancer-related targets, and the intersecting targets are the potential targets of garlic action in gastric cancer.

2.8.3. GO function and KEGG pathway enrichment analysis

Based on the R language package "ClusterProfiler", we conducted GO and KEGG enrichment analysis. We obtained the main functions and pathways enriched for potential targets. The GO enrichment analyses include three main aspects: biological processes, cellular components, and molecular functions.

2.8.4. Protein Interaction Network Construction

Import the potential target into STRING, select the organism as "Homo sapiens", set the required score as "highest confidence ( ≥ 0.700)", and then construct the protein interaction network [37]. We exported the protein interactions file in TSV format and performed topological analyses using the CytoNCA plug-in of Cytoscape 3.8.0 software to screen the core targets in the protein interaction network [38].

2.8.5. Construction of drug-component-disease-target-pathway network diagram

Cytoscape 3.8.0 was used to construct a "drug-component-disease-target-pathway network plot", and the network plot was topologically analyzed to screen the core components in garlic.

2.8.6. Molecular docking

The three-dimensional (3D) structure of the core target was downloaded from the PDB database [39] and we acquired the three-dimensional structures of the core components of garlic from the PubChem database (https://pubchem.ncbi.nlm.nih.gov/). We imported the core targets into PyMOL software (Version 2.5.2) to delete water molecules and ligands. Then, the target was imported into Autodock tool 1.5.6 for adding hydrogen bonds, calculating total charge, and saving as a receptor. We used Autodock-vina to molecularly dock the receptor and ligand and then visualized it using PyMOL.

**3. Results**

*3.1. Selection of Instrumental Variables (IVs)*

19 SNPs associated with garlic intake and 14 SNPs associated with onion intake were obtained. To remove confounding, we searched for garlic and onion related SNPs in LDtrait and IEU QpenGWAS project respectively. In the end, we still retained 14 SNPs associated with onion intake. But for garlic intake, we found that rs56269735 was associated with BMI, so we removed this SNP. In addition, all IVs had F statistic values greater than 10. The details of all instrumental variables are shown in Supplementary Table 2.

*3.2. Mendelian randomization results*

Analyzed by a two-sample MR method based on IVW, we obtained the following results: Onion intake is not causally correlated with digestive system cancers (p>0.05). Garlic intake is not causally associated with esophageal, liver, pancreatic and colorectal cancers (esophageal cancer: OR= 1.760,95%CI:0.151-20.531, p=0.652; liver cancer: OR= 2.545,95%CI:0.236-27.437, p=0.441; pancreatic cancer: OR= 0.712,95%CI:0.298-1.700, p=0.445; colorectal cancer: OR= 1.039,95%CI:0.773-1.396, p=0.800). Garlic intake was causally correlated with gastric cancer (OR= 0.691, 95% CI 0.484-0.986, p=0.041). Each standard deviation increase in garlic intake was associated with a 30.9% decrease in the incidence of gastric cancer. Figure 2 illustrates the results of the Mendelian randomization. Figure 3 demonstrates a scatter plot of the correlation between allium vegetable intake and digestive system cancers.

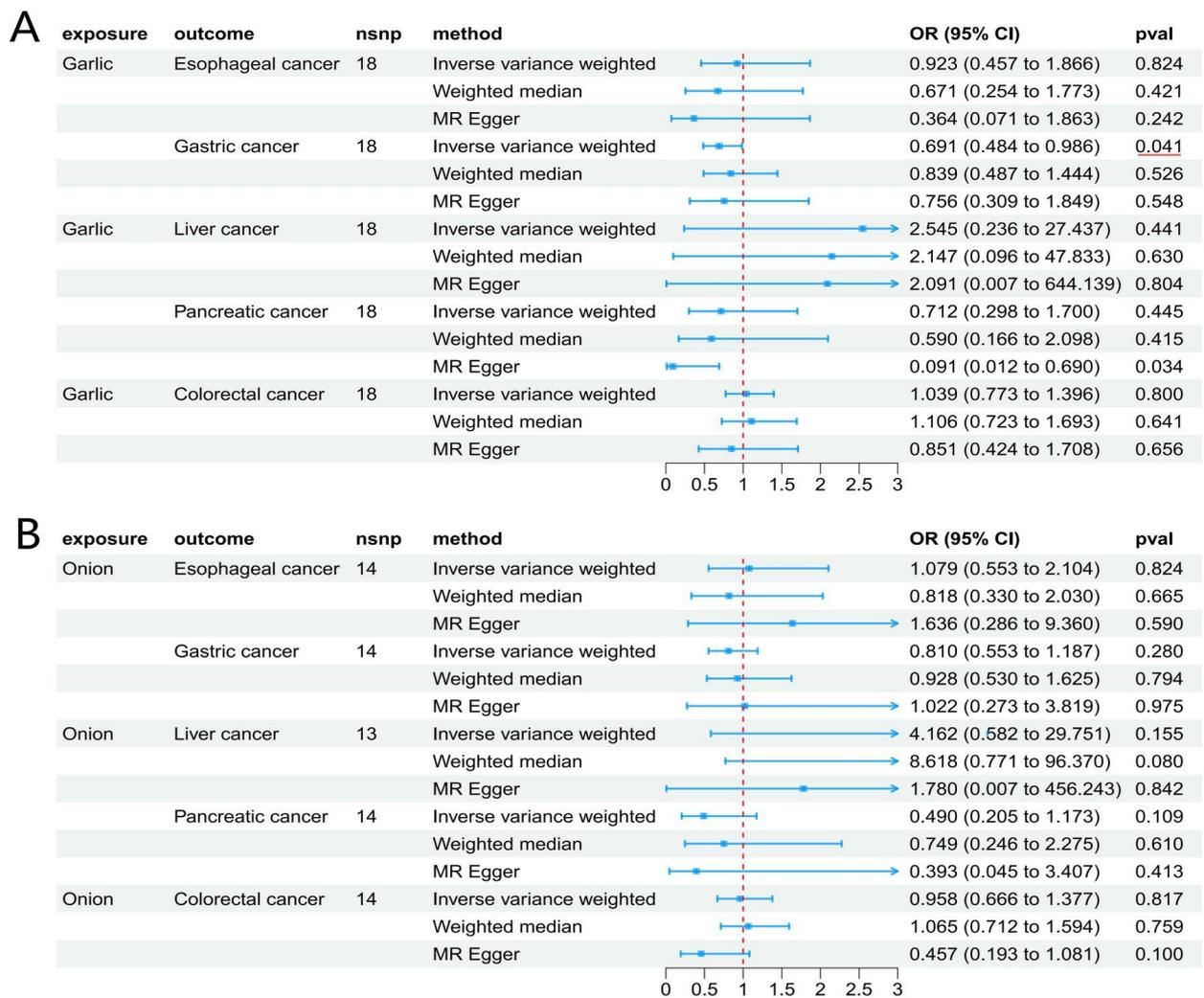

**Figure 2.** Forest plot showing the causal relationship between allium vegetables intake and digestive system cancers risk (A); Causal association between garlic intake and risk of digestive system cancers; Causal association between onion intake and risk of digestive system cancers (B); OR: odds ratio.

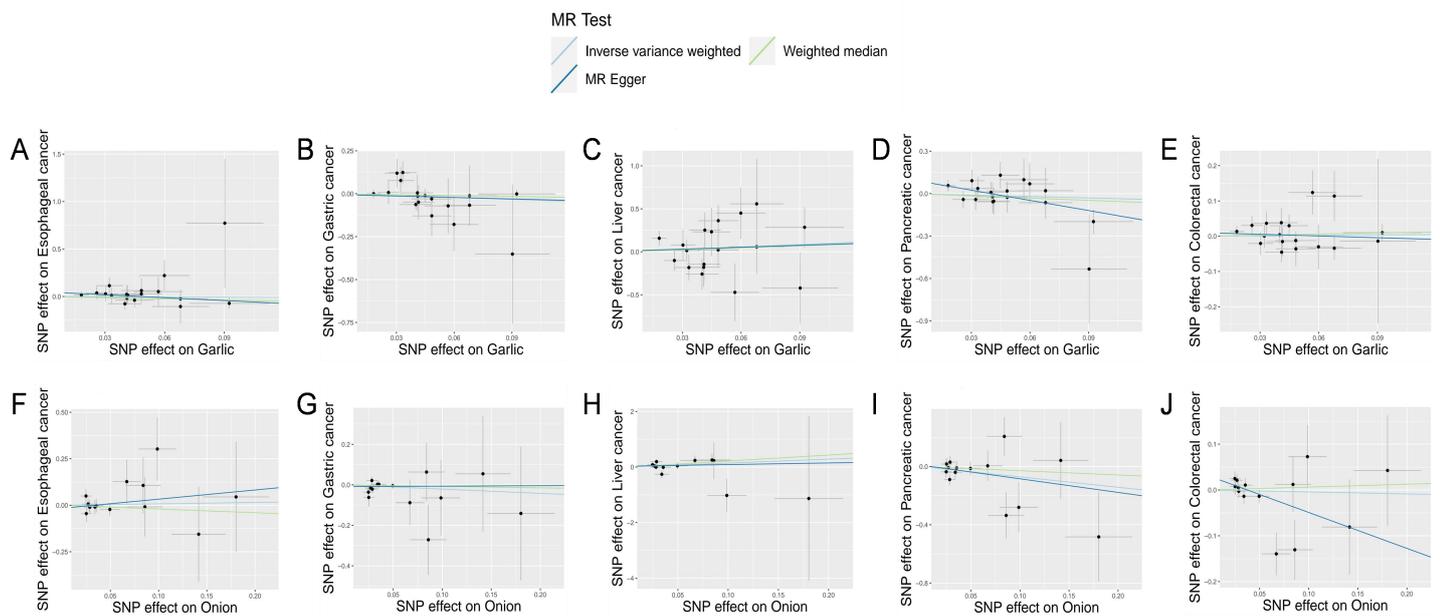

**Figure 3.** Scatter plots of correlation between allium vegetables intake and digestive system cancers using different MR methods. Scatterplot of correlation between garlic intake and esophageal cancer (A), gastric cancer (B), liver cancer;(C), pancreatic cancer (D), colorectal cancer (E); Scatterplot of correlation between onion intake and esophageal (F), gastric cancer (G), liver cancer (H), pancreatic cancer (I), and colorectal cancer (J).

*3.3. Sensitivity analyses*

In the sensitivity analysis stage, we performed the following analyses: We used Cochrane's Q-test to test for heterogeneity in the instrumental variables, and the results showed that no heterogeneity existed. MR-PRESSO analysis did not detect outliers, and the global P value was >0.05. MR-Egger intercepts did not significantly deviate from 0, which indicates that there was no significant horizontal pleiotropy. The leave-one-out method analysis proves that the results of this study are not driven by a single instrumental variable (Figure 4). In summary, the IVW results in the study were sufficiently reliable. The results of the sensitivity analysis are presented in Supplementary Table 3.

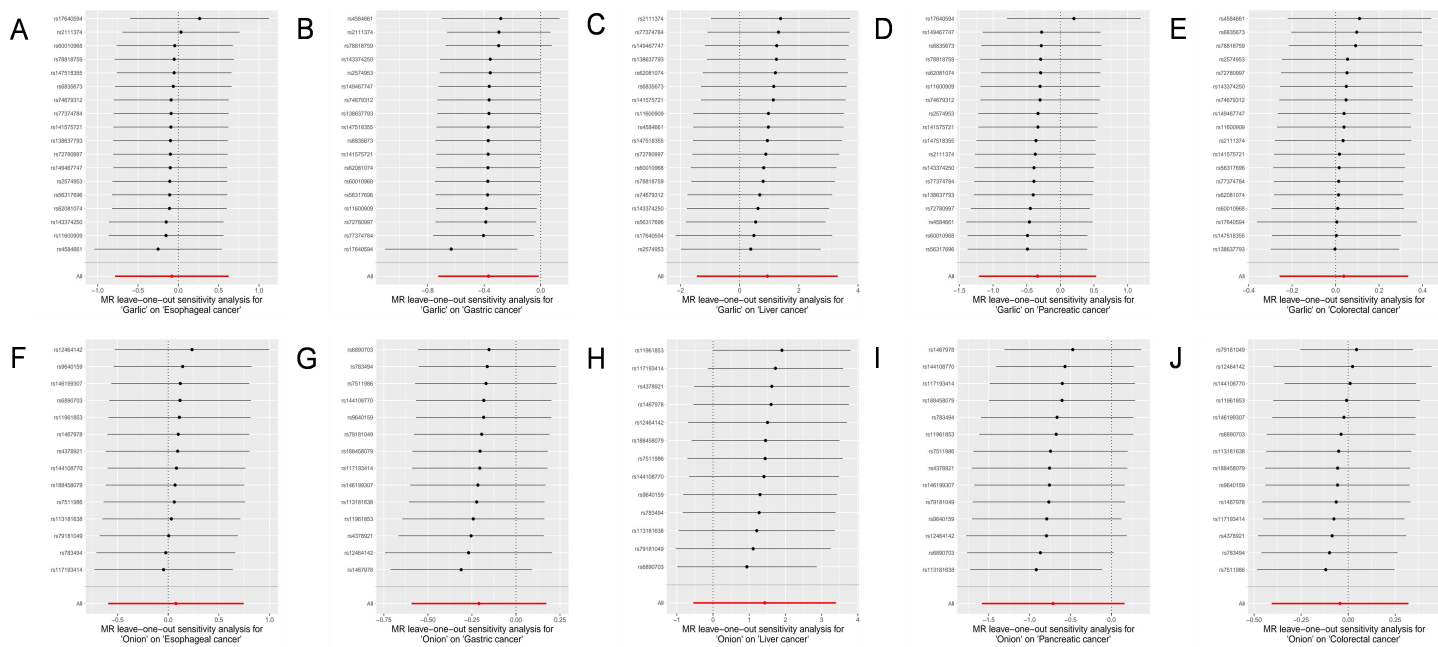

**Figure 4.** Leave-one-out method to analyze the driving role of individual SNPs in the causal relationship between allium vegetables intake and digestive system cancer risk.

*3.4. Results of multivariable Mendelian randomization analysis*

Inverse variance weighted (IVW) analysis suggests that there was still a causal relationship between garlic intake and gastric cancer (OR= 0.507, 95% CI: 0.291-0.883, P=0.017) after adjusting for smoking and alcohol consumption. There were no causal associations between garlic intake and esophageal cancer (p=0.651), liver cancer (p=0.647), pancreatic cancer (p= 0.858), colorectal cancer (p=0.184) were not causally related. In addition, there is still no causal relationship between onion intake and digestive system cancers. Table 2 demonstrates the results of multivariable Mendelian randomization analyses after adjusting for smoking and alcohol consumption.

Table 2. Results of a multivariate mendelian randomization analysis of allium vegetables intake and digestive system cancers after adjusting for smoking and alcohol consumption.

| Exposure | Outcome | Method | P value | OR(95% CI) | Q.val | P.int |
|---|---|---|---|---|---|---|
| Garlic | Esophageal cancer | IVW | 0.519 | 0.292(0.007 to 12.329) | 0.867 | - |
|  |  | Weighted Median | 0.485 | 2.687(0.167 to 43.139) | - | - |
|  |  | MR Egger | 0.570 | 3.070(0.064 to 147.125) | 0.814 | 0.952 |
| Garlic | Gastric cancer | IVW | 0.017 | 0.507(0.291 to 0.883) | 0.753 | - |
|  |  | Weighted Median | 0.695 | 0.938(0.682 to 1.290) | - | - |
|  |  | MR Egger | 0.309 | 0.758(0.444 to 1.294) | 0.716 | 0.514 |
| Garlic | Liver cancer | IVW | 0.647 | 2.479(0.051 to 121.204) | 0.265 | - |
|  |  | Weighted Median | 0.922 | 0.931(0.222 to 3.912) | - | - |
|  |  | MR Egger | 0.827 | 1.245(0.174 to 8.903) | 0.213 | 0.697 |
| Garlic | Pancreatic cancer | IVW | 0.858 | 1.140(0.271 to 4.802) | 0.823 | - |
|  |  | Weighted Median | 0.634 | 1.180(0.597 to 2.332) | - | - |
|  |  | MR Egger | 0.926 | 1.040(0.462 to 2.340) | 0.817 | 0.393 |
| Garlic | Colorectal cancer | IVW | 0.184 | 0.718(0.440 to 1.171) | 0.371 | - |
|  |  | Weighted Median | 0.107 | 1.222(0.957 to 1.561) | - | - |
|  |  | MR Egger | 0.566 | 1.107(0.783 to 1.565) | 0.296 | 0.940 |
| Onion | Esophageal cancer | IVW | 0.242 | 6.351(0.287 to 104.353) | 0.420 | - |
|  |  | Weighted Median | 0.338 | 3.810(0.247 to 58.683) | - | - |
|  |  | MR Egger | 0.788 | 1.450(0.967 to 21.740) | 0.414 | 0.327 |
| Onion | Gastric cancer | IVW | 0.418 | 0.840(0.550 to 1.281) | 0.713 | - |
|  |  | Weighted Median | 0.779 | 0.957(0.704 to 1.301) | - | - |
|  |  | MR Egger | 0.534 | 0.875(0.575 to 1.332) | 0.653 | 0.660 |
| Onion | Liver cancer | IVW | 0.640 | 2.032(0.104 to 39.674) | 0.217 | - |
|  |  | Weighted Median | 0.917 | 1.075(0.275 to 4.208) | - | - |
|  |  | MR Egger | 0.899 | 0.877(0.116 to 6.613) | 0.171 | 0.742 |
| Onion | Pancreatic cancer | IVW | 0.870 | 0.915(0.317 to 2.642) | 0.914 | - |
|  |  | Weighted Median | 0.714 | 1.123(0.603 to 2.091) | - | - |
|  |  | MR Egger | 0.771 | 1.123(0.515 to 2.451) | 0.882 | 0.652 |
| Onion | Colorectal cancer | IVW | 0.813 | 0.952(0.633 to 1.432) | 0.184 | - |
|  |  | Weighted Median | 0.241 | 1.157(0.906 to 1.477) | - | - |
|  |  | MR Egger | 0.398 | 1.162(0.821 to 1.645) | 0.151 | 0.599 |

*3.5. Network pharmacology and molecular docking*

3.5.1. Acquisition of garlic active ingredients and related targets

Based on the HERB database, 79 active ingredients in garlic were screened. After obtaining SMILES numbers in the PubChem database, we obtained 553 garlic-related targets by predicting garlic-related targets in Swiss Target Prediction.

3.5.2. Acquisition of targets related to gastric cancer

The Genecards and OMIM databases were searched with the keyword "gastric cancer", and a total of 1541 targets related to gastric cancer were found. We took the intersection of garlic and gastric cancer targets and obtained 149 potential targets of garlic for gastric cancer.

3.5.3. GO and KEGG enrichment analysis

We conducted GO and KEGG enrichment analysis of potential targets. GO enrichment analysis elaborated the functions involved in effective targets at three dimensions: biological process, cellular component, and molecular function. We found that at the biological process level, intersecting targets are mainly enriched in positive regulation of transferase activity, response to oxidative stress, and positive regulation of kinase activities and other functions. At the cellular component level, intersecting targets are mainly enriched in focal adhesions, cell-substrate junctions, transferase complexes and other functions. At the molecular functional level, intersecting targets are mainly enriched in protein serine/threonine kinase activity, protein tyrosine kinase activity, protein serine kinase activity and other functions. KEGG enrichment analysis indicated that the cross-targets were enriched mainly in PI3K-AKT signaling pathway, FoxO signaling pathway, etc. The Figure 5 shows the results of GO and KEGG enrichment analysis.

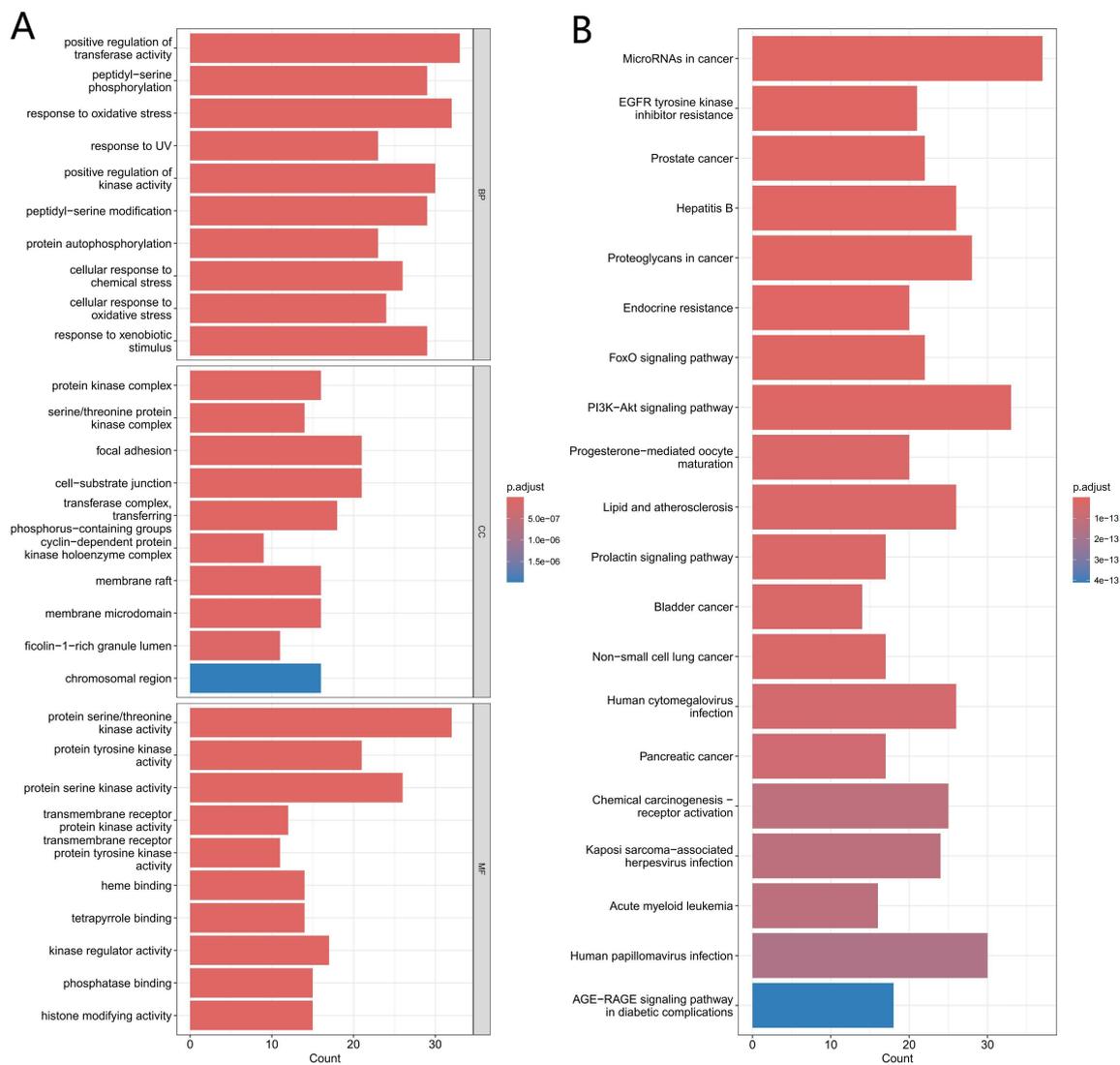

**Figure 5.** Enrichment analysis of garlic for gastric cancer. (A) GO enrichment analysis; (B) KEGG enrichment analysis.

3.5.4. Construction of protein interaction network

We constructed a protein interaction network diagram (Figure 6). The protein interaction network was topologically analyzed, and its core targets were HSP90AA1, STAT3, epidermal growth factor receptor, SRC, AKT1, and BCL2, which may have an important impact on the prevention of gastric cancer by garlic. Table 3 demonstrates the topological traits of the core targets. Table 3 demonstrates the topological traits of the core targets.

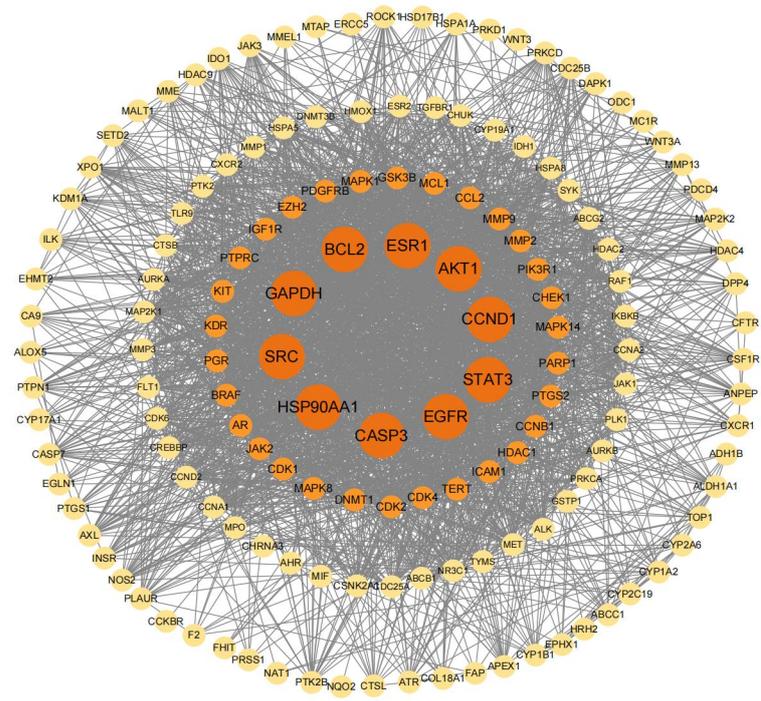

**Figure 6.** PPI network diagram of intersecting targets. Circles represent proteins, ranging in color from orange to pale yellow, with darker colors representing more important proteins, and lines between proteins representing interactions between proteins.

**Table 3.** Topological properties of the core target.

| NO | Gene | Targets | Degree | Betweenness Centrality | Closeness Centrality |
|----|------|---------|--------|------------------------|----------------------|
| 1 | HSP90AA1 | Heat Shock Protein 90 Alpha Family Class A Member 1 | 53 | 0.150160936 | 0.572016461 |
| 2 | STAT3 | Signal Transducer and Activator of Transcription 3 | 52 | 0.101347483 | 0.567346939 |
| 3 | EGFR | Epidermal Growth Factor Receptor | 46 | 0.100813724 | 0.567346939 |
| 4 | SRC | Proto-Oncogene, Non-Receptor Tyrosine Kinase | 45 | 0.058939037 | 0.549407115 |
| 5 | AKT1 | Serine/Threonine Kinase 1 | 44 | 0.062501444 | 0.562753036 |
| 6 | BCL2 | Apoptosis Regulator | 41 | 0.063583881 | 0.545098039 |
| 7 | CCND1 | Cyclin D1 | 36 | 0.043065223 | 0.534615385 |
| 8 | ESR1 | Estrogen Receptor 1 | 34 | 0.037958241 | 0.518656716 |
| 9 | CASP3 | Caspase 3 | 31 | 0.0297224 | 0.516728625 |
| 10 | GAPDH | Glyceraldehyde-3-Phosphate Dehydrogenase | 31 | 0.064870613 | 0.540856031 |

3.5.5. Construction of drug-component-disease-target-pathway network diagram

Cytoscacpe was used to construct the drug-component-disease-target-pathway network diagram (Figure 7). We analyzed all the nodes in the network diagram topologically using Network Analyzer and filtered the core compounds of garlic based on degree. We found that core compounds such as kaempferol and apigenin act on multiple targets, and they may be the main active substances in garlic for gastric cancer prevention. The topological properties of the core components in garlic are shown in Table 4.

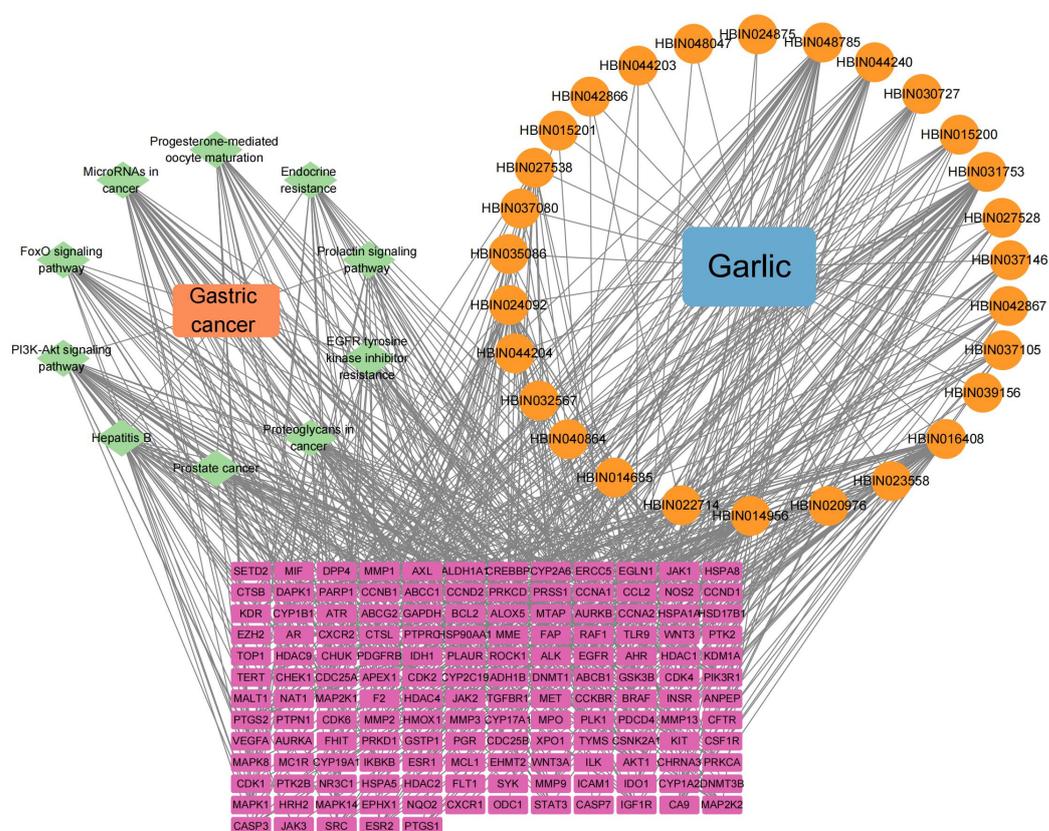

**Figure 7.** The drug-component-disease-target-pathway network diagram. Purple rectangles are cross-targets, orange circles are garlic active compounds, and green diamonds are pathways associated with intersecting targets.

**Table 4.** Topological properties of the core components of garlic.

| NO | Herb ID | Molecule Name | Degree | Betweenness Centrality | Closeness Centrality |
| --- | --- | --- | --- | --- | --- |
| 1 | HBIN031753 | Kaempferol | 44 | 0.110026898 | 0.428246014 |
| 2 | HBIN016408 | Apigenin | 44 | 0.110650024 | 0.428246014 |
| 3 | HBIN014956 | Ajoene | 36 | 0.10495414 | 0.413186813 |
| 4 | HBIN024092 | Dihydrosamidin | 33 | 0.100352882 | 0.407809111 |
| 5 | HBIN048785 | Z-4,9-Diene-2,3,7-trithiadeca-7-oxide | 31 | 0.098301617 | 0.404301075 |
| 6 | HBIN044240 | Ethylparaben | 22 | 0.082125383 | 0.389233954 |
| 7 | HBIN014685 | Adenosine | 19 | 0.101738979 | 0.384458078 |
| 8 | HBIN027538 | [(1E)-2,6-dimethylhepta-1,5-dienyl] acetate | 17 | 0.050642111 | 0.37979798 |
| 9 | HBIN030727 | Isoeugenitol | 15 | 0.015050481 | 0.375249501 |
| 10 | HBIN037105 | N-Methylmescaline | 14 | 0.063331146 | 0.376753507 |
| 11 | HBIN015200 | Allicin | 12 | 0.009817622 | 0.36935167 |
| 12 | HBIN042867 | s-Allyl mercaptocysteine | 11 | 0.029881835 | 0.359464627 |
| 13 | HBIN023558 | Diallyl disulfide | 10 | 0.006218606 | 0.355387524 |
| 14 | HBIN020976 | Neral | 9 | 0.037277129 | 0.365048544 |
| 15 | HBIN037146 | Dimethyltryptamine | 9 | 0.029474214 | 0.363636364 |

3.5.6. Molecular docking

It is commonly accepted that a binding energy of -4.25 kcal/mol demonstrates binding activity between the ligand molecule and the receptor protein; a binding energy of -5.0 kcal/mol indicates good binding activity between the two; and the binding energy of -7.0 kcal/mol indicates a robust binding activity between the two [40]. We molecularly docked the 15 core components screened with ten core targets. Some of the macromolecular docking pattern diagrams are shown in Figure 8.

Binding energy is used to assess the stability of binding between core components and core targets. The lower the binding energy, the more stable the binding and the more easily the compound binds to the target. We found that the core compounds Kaempferol, Apigenin and Dihydrosamidin have binding energies less than -5 kcal/mol with all core targets, which proves that Kaempferol, Apigenin have good binding activities with core targets. The binding energies of all core compounds are less than -2 kcal/mol with core targets, proving that core compounds and core targets can bind independently. Figure 9 illustrates a heat map of the binding energy of all the core compounds to the core targets.

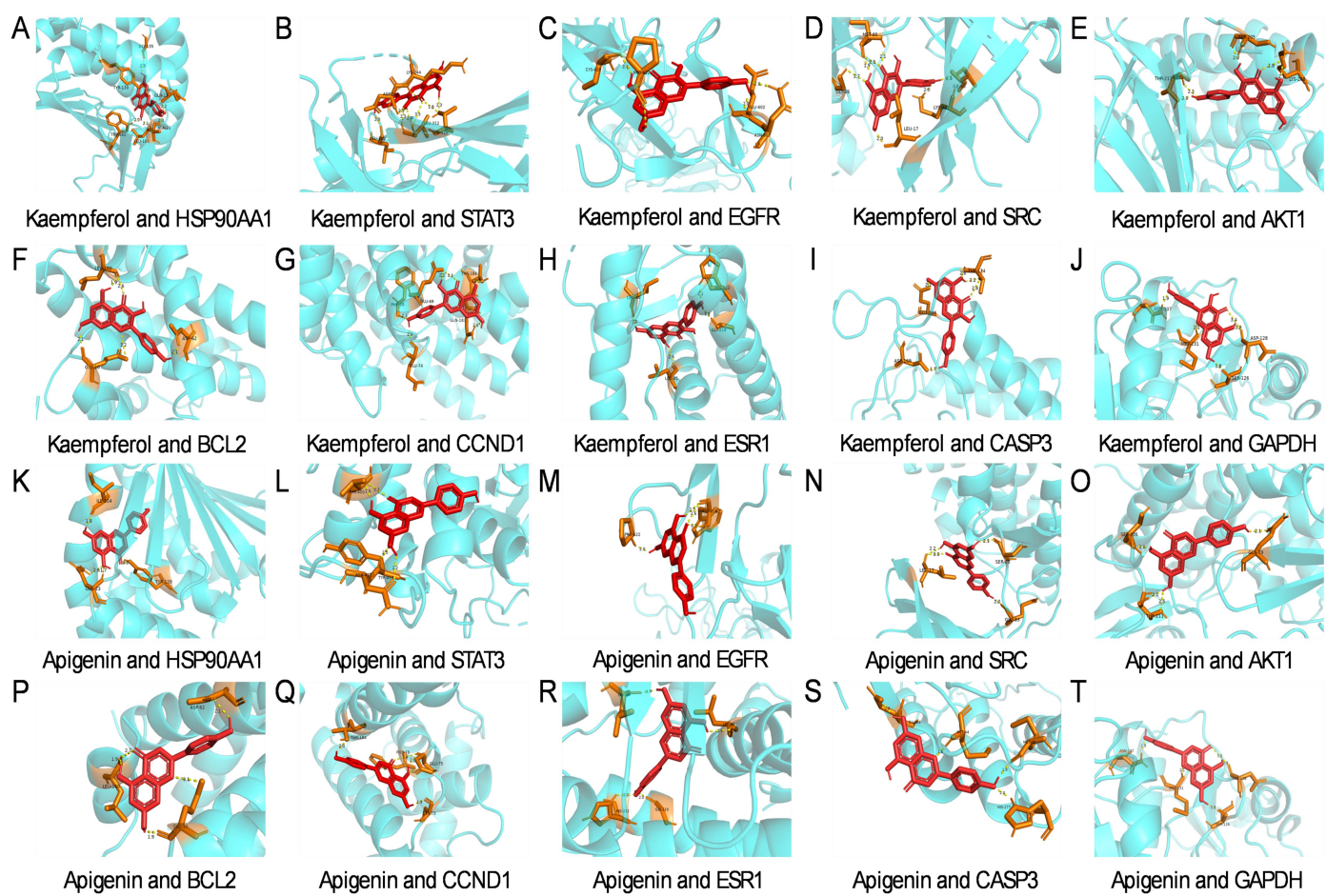

**Figure 8.** Macromolecular docking pattern diagram. Docking plots of kaempferol with HSP90AA1 (A), STAT3 (B), EGFR (C), SRC (D), AKT1 (E), BCL2 (F), CCND1 (G), ESR1 (H), CASP3 (I), GAPDH (J); Docking plots of apigenin with HSP90AA1 (K), STAT3(L), EGFR (M), SRC (N), AKT1 (O), BCL2 (P), CCND1 (Q), ESR1 (R), CASP3 (S), and GAPDH (T).

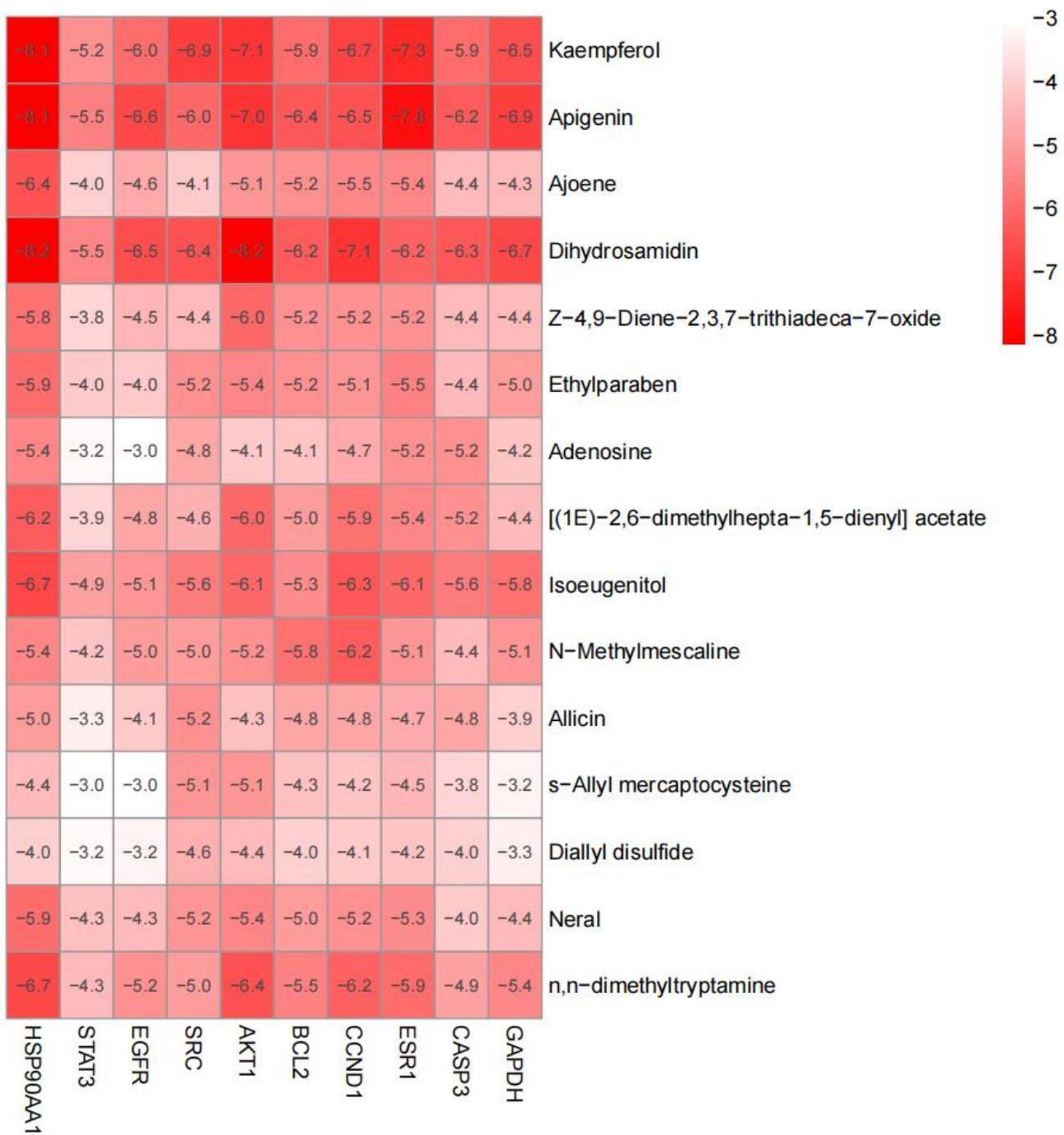

**Figure 9.** Binding energy heatmap for molecular docking of active compounds to core targets. The color ranges from white to red, with the darker color representing the compound's lower binding energy to the target and the easier the compound binds to the target.

## 4. Discussion

In the study, we evaluated the causal association between onion and garlic vegetable intake and digestive system cancers by magnetic resonance analysis, which is the first time this relationship has been explored from a genetic epidemiological perspective. We selected common digestive system cancers, including esophageal, gastric, liver, pancreatic, and colorectal cancers, and our analyses showed a negative correlation between garlic intake and gastric cancer risk. The reliability of the conclusion was demonstrated through sensitivity analysis and multivariable Mendelian randomization analysis. We also performed network pharmacology and molecular docking approaches to explore the molecular mechanisms between garlic and gastric cancer.

Our study shows consistency with many previous studies. A meta-analysis based on case-control and cohort studies demonstrated that allium vegetable intake can reduce gastric cancer risk (OR, 0.54; 95% CI: 0.43-0.65) [41]. In addition, another study separately analyzed the causal relationship between garlic and gastric cancer and showed that garlic intake was negatively associated with gastric cancer (OR=0.51,95% CI=0.44-0.57) [42]. A meta-analysis based on the relationship between garlic consumption and cancer prevention also demonstrated that high intake of garlic was a protective factor against gastric cancer [16]. For esophageal cancer, the European Prospective Investigation into Cancer and Nutrition indicated that garlic and onion intake was not correlated with esophageal cancer (calibrated HR 1.54, 95% CI 0.72-3.28 per 10 g increase) [14]. However, our study is also inconsistent with some previous studies on the relationship between allium vegetable intake and digestive system cancers. The results of a cohort study in the United States suggest that consuming large amounts of garlic does not reduce the risk of gastric cancer [12]. Moreover, a population-based case-control study in eastern China suggested that consuming garlic twice a week or more may have a preventive influence on liver cancer (95% CI: 0.62-0.96) [13]. For colorectal cancer, a meta-analysis of studies indicated that high intake of garlic may be protective against colorectal cancer [16].

Based on our results, various biological mechanisms can account for the negative correlation between garlic intake and gastric cancer risk. We found that the main components of garlic that have therapeutic effects on gastric cancer include kaempferol, apigenin, and allicin. Kaempferol can inhibit gastric cancer cell proliferation by decreasing the Cdk1/cyclin B kinase complex, leading to G2 cell blockade [43]. It can also down-regulate the expression of key proteins p-AKT, p-ERK and cox-2 in the survival pathway and activate the endogenous cell death pathway to induce apoptosis in gastric cancer cell [44]. Apigenin suppresses cell proliferation and induces apoptosis in gastric cancer cells by activating the Akt/Bad/Bcl2/Bax axis of the mitochondrial apoptotic pathway, leading to the restriction of cell proliferation [45]. In addition, apigenin induced autophagy in gastric cancer cells. Its mechanism may be related to inhibiting the activities of p-Akt and p-mTOR, key molecules in the PI3K/Akt/mTOR signaling pathway [46]. Allicin is a sulfur-containing natural compound that induces gastric cancer cell apoptosis by increasing the expression of p38 and cleaved caspase-3 through the p38 MAPK/caspase-3 signaling pathway [47]. Most of the actions of these compounds are accomplished through the PI3K-AKT pathway, which is similar to the findings of this study.

In addition, potential garlic targets for preventing gastric cancer include HSP90AA1, STAT3, EGFR, SRC, AKT1, BCL2, CCND1, ESR1, and GADPH. The HSP90AA1 gene encodes HSP90a, which plays an essential function in stabilizing oncogenic proteins and promoting cancer cell survival. HSP90AA1 is critical in DNA damage, cell cycle, and gene expression. Therefore, HSP90AA1 may be a potential core target for treating gastric cancer [48-50]. BCL2, a protein encoded by the BCL2 gene, which plays a regulatory role in the proliferation and apoptosis of tumor cells. The study indicated that knocking down the BCL2 gene induced intrinsic apoptotic pathways associated with mitochondrial dysfunction and cysteine asparaginase activation in gastric cancer cells [51]. SRC is the first oncogenic gene discovered, and some studies have shown that inhibition of SRC protein can inhibit gastric cancer cell proliferation and promote apoptosis [52]. We suggested that core compounds in garlic, such as quercetin and apigenin, may play a role in preventing gastric cancer by acting on the above targets.

Our study has multiple advantages. The MR approach reduces the risk of confounding and reverse causation, which is the greatest strength of the Mendelian randomization analysis approach. In our choice of instrumental variables: firstly, we ensured that IVs were strongly correlated with exposure (P<5E-06); secondly, that IVs were not affected by confounders such as the environment, and we removed shared confounders through both the LDtrait and IEU Open GWAS project databases; ultimately, that IVs could only influence outcome through exposure factors, with no horizontal pleiotropy. We also made our results more robust through multiple MR methods and sensitivity analyses. After adjusting for smoking and alcohol consumption, Multivariable Mendelian randomization provided a more valid estimate of the causal relationship between allium vegetables and cancers of the digestive system.

Nevertheless, there are limitations to our study. Firstly, there may be intrinsic variability in the intake of allium vegetables in different populations, e.g., the way garlic and onions are cooked, and whether the allium vegetables are used raw or cooked may affect the concentration of the compounds. Additionally, it has been shown that organosulfur compounds in raw garlic are more digestible than cooked garlic [53]. Secondly, the IVs used in the study were derived from UK biobank's 24-hour dietary recalls, and participants may have inadvertently exaggerated or underestimated their allium vegetable intake over the past 24 hours, resulting in recall and response bias that may have impacted our study. In the future, precise measurement of garlic intake achieved by more efficient methods may reduce this bias. Third, MR methods cannot exclude all potential confounders, and some SNP-related traits may be associated with digestive system cancers but have not been reported in the literature. Finally, our study was limited to European populations, and there may be significant differences in lifestyle and dietary habits between ethnic groups [54], which may interact with garlic intake to influence digestive system cancer risk. Despite the above limitations, the present study is one of the more comprehensive MR studies evaluating the causal relationship between Allium vegetable intake and digestive system cancers. The limitations described above indicate the need for further exploration in this field. Future studies could replicate our findings in other races and further explore the molecular mechanisms of action of garlic intake in preventing gastric cancer.

5.Conclusions

In the study, we assessed the causal relationship between intake of allium vegetables and digestive system cancers. We conclude that increased intake of garlic may have a protective role against gastric cancer. Through network pharmacology and molecular docking, we explored the molecular mechanism of garlic action in gastric cancer and screened the core targets and compounds of garlic action in gastric cancer. Adjusting garlic intake may be an effective measure for preventing gastric cancer.

**Author Contributions**: Conceptualization, S.L.; data curation, J.L., Y.M., Y.X., formal analysis, S.L.; funding acquisition, J.L. and Y.M.; resources, J.L., Y.M., Y.X.; software, S.L..; supervision, Q.X.; validation, S.L. and Y.L.; visualization, S.L.; writing—original draft, S.L.; writing—review and editing, J.L. and Y.X. All authors have read and agreed to the published version of the manuscript.

**Funding:** This work was supported by College Students' Innovative Entrepreneurial Training Plan Program, item number S202410760052.

**Institutional Review Board Statement:** Not applicable.

**Informed Consent Statement:** Not applicable.